\documentclass[%
reprint,
superscriptaddress,
 amsmath,amssymb,
aps,
 prb,
]{revtex4-2}

\usepackage[unicode=true,colorlinks=true,citecolor=blue,urlcolor=blue]{hyperref}
\usepackage[english]{babel} 
\usepackage{amssymb}
\usepackage{amsmath}
\usepackage{txfonts}
\usepackage{mathdots}
\usepackage[classicReIm]{kpfonts}
\usepackage{graphicx}
\usepackage[dvipsnames]{xcolor}
\def\arrvline{\hfil\kern\arraycolsep\vline\kern-\arraycolsep\hfilneg}
\usepackage{dcolumn}
\usepackage{bm}
\usepackage{multirow}
\usepackage[normalem]{ulem}
\usepackage{tabularx}
\usepackage{comment}

\newcommand{\e}{\mathrm{e}}
\renewcommand{\d}{\mathrm d}
\renewcommand{\Re}{\mathop{\rm Re}}
\let\ifr\i
\renewcommand{\i}{{\rm i}}

\begin{document}
\preprint{APS/123-QED}
\title{Spin noise of localized electrons in CdTe/CdMgTe quantum well}
\author{A.~L.~Zibinskiy}
\affiliation{Ioffe Institute, 194021 St. Petersburg, Russia}
\author{S.~Cronenberger}
\affiliation{Laboratoire Charles Coulomb, UMR 5221 CNRS/Universit\'{e} de Montpellier, France}
\author{B.~Gribakin}
\affiliation{Laboratoire Charles Coulomb, UMR 5221 CNRS/Universit\'{e} de Montpellier, France}
\author{R.~Baye}
\affiliation{Laboratoire Charles Coulomb, UMR 5221 CNRS/Universit\'{e} de Montpellier, France}
\author{D.~Scalbert}
\affiliation{Laboratoire Charles Coulomb, UMR 5221 CNRS/Universit\'{e} de Montpellier, France}
\author{R.~André}
\affiliation{Universit\'e Grenoble Alpes, CNRS, Institut N{\'{e}}el, 38000 Grenoble, France}
\author{D.~S.~Smirnov}
\affiliation{Ioffe Institute, 194021 St. Petersburg, Russia}
\author{M.~Vladimirova}
\affiliation{Laboratoire Charles Coulomb, UMR 5221 CNRS/Universit\'{e} de Montpellier, France}

\begin{abstract}

  The spin dynamics of localized electrons in bulk semiconductors is governed by the interplay of effective nuclear field fluctuations, spin exchange between electrons, and spin transitions into the conduction band. Using spin noise spectroscopy, we reveal this interplay for donor-bound electrons in a CdTe/CdMgTe quantum well and spectrally separate electron spin relaxation and dephasing in zero magnetic field.
  We identify a specific regime of the electron spin dynamics, where temperature-induced activation of spin-independent hopping leads to a monotonic acceleration of electron spin relaxation.  This behavior contrasts with bulk CdTe crystals, where the motional narrowing effect is observed. We attribute this difference to the significantly larger inhomogeneous broadening of the donor-related trion resonance in our quantum well compared to bulk samples.
  The theoretical analysis of the spin noise power and the strength of the spin exchange interaction provides the estimation of the donor concentration in our unintentionally doped structure.
\end{abstract}

\date{\today}
\maketitle
\section{Introduction}

Localised 
charge carriers in direct band gap semiconductors attract considerable attention in view of their potential implementation in quantum technologies. The interest is driven mainly by the long spin relaxation times of localized {electrons} and their {peculiar} spin dynamics~\cite{KKavokin-review,Gribakin2025, book_Glazov}. In this context, spin properties of donor-bound electrons in bulk semiconductors \cite{Dzhioev2002,belykh_electron_2017,Sterin2022,Gribakin2025}, as well as electrons confined in quantum wells {(QWs)~}\cite{zhukov_spin_2009,Garcia-Arellano2019,garcia-arellano_longitudinal_2020,kosarev_microscopic_2019} and dots \cite{merkulov02,Nonresonant_nonequilibrium,krebs_hyperfine_2008,Glazov_effect_2010,crooker_spin_2010,tartakovskii_nuclear_2012}, has been studied extensively.

It has been established, both theoretically and experimentally, that at low electron densities and small magnetic fields, three main parameters control electron spin properties. The first {one} is the strength of the hyperfine interaction with the slowly fluctuating nuclear spins {located} within electron orbit~\cite{book_Glazov}. It is characterized by the typical  frequency of electron spin precession $\delta_e$, corresponding to the effective nuclear magnetic field. It determines the reversible electron spin dephasing time $T_2^*\sim\delta_e^{-1}$.

The second {parameter} is often called the electron spin correlation time, $\tau_c$~\cite{KKavokin-review}. For donor-bound electrons at low temperatures, $T\approx 4$~K, $\tau_c$ is {often} limited by the spin exchange with electrons localized on neighboring  donors~\cite{kkavokin01,shklovskii2006,lyubinskiy07}. Typically, at moderate donor densities, spin exchange is weak, so that $\delta_e \tau_c {\gg}1 $. In this {regime,} the longitudinal spin relaxation is limited by the correlation time, $T_1{\sim}\tau_c$.

When the temperature increases, frequent jumps of donor-bound electrons back and forth in the conduction band lead to the suppression of the nuclei induced spin dephasing~\cite{Glazov_hopping,PRC,SmirovKavokin2025} via motional narrowing  effect~\cite{Croneberger2021,Sterin2022,Gribakin2025}. Then, at higher temperatures, electrons delocalize,  and other mechanisms, such as spin-orbit interaction and interaction with phonons,  become dominant~\cite{Spin2017}.

\begin{figure*}
\includegraphics[width=6in]{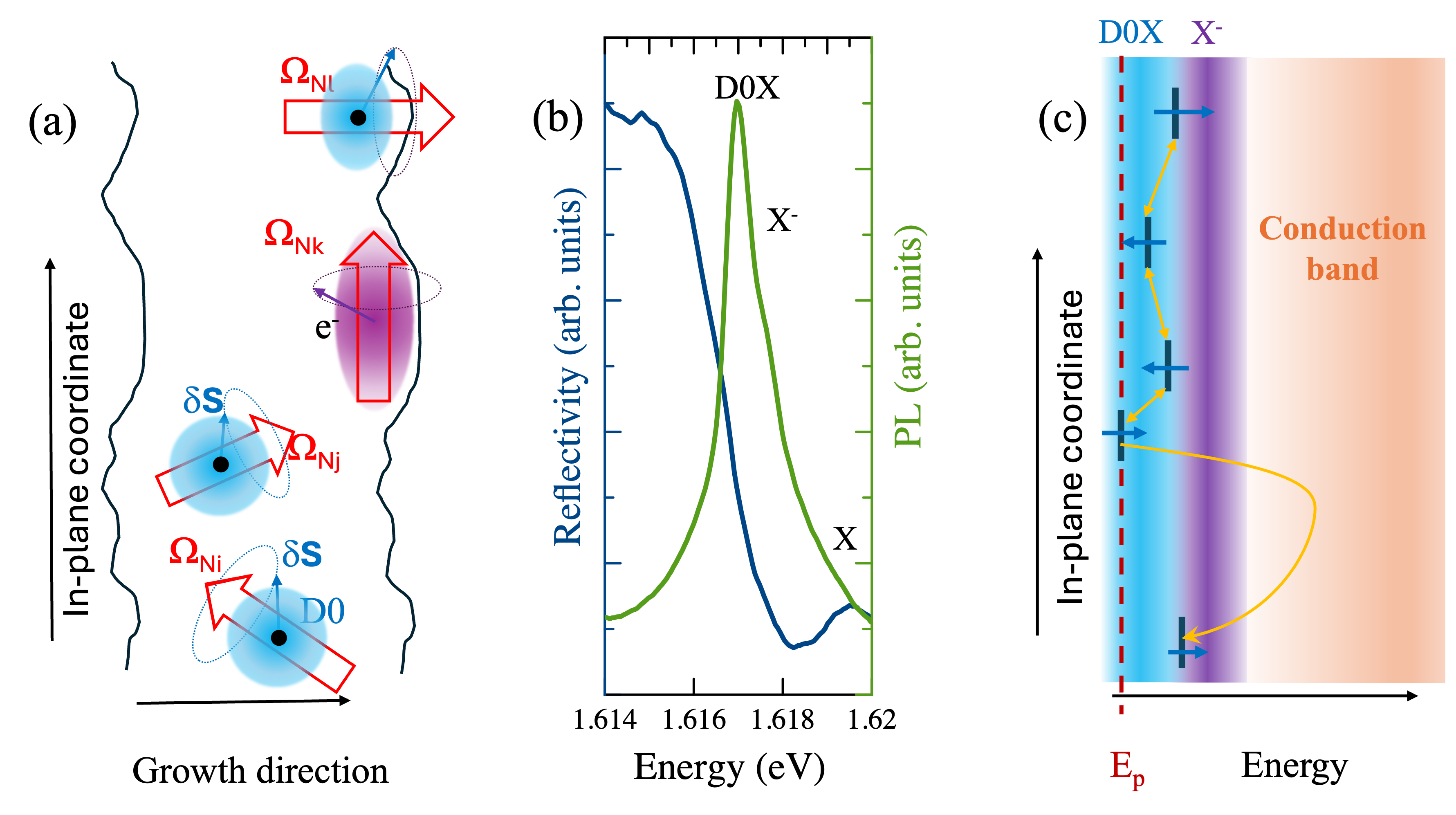}
\caption{ (a) PL and reflectivity spectra of the QW, with donor-bound exciton, negative trion, and exciton identified. 
(b)  Artistic view of various spin species localized in the QW: donor-bound electrons in the QW center and close to the interface (blue), as well as electrons localized at the interface roughness (magenta). The latter, as well as the electrons localized on donors close to the QW edge have the wave-function different from the spherical one, characterized by the bulk Bohr radius. Each of the spin fluctuations $\delta\bm S$ precesses around an effective field $\bm\Omega_{N,i}$ of the underlying nuclei, .
(c) Sketch of the excited states energy levels. Red dashed line shows the probe laser energy, $E_p$, tuned to the bottom of the D0X energy distribution (blue band), well below $X^-$ band (magenta).  Yellow lines schematize the mechanisms leading to the spin loss within the probed energy band: electron jumps via conduction band (curved arrow) and electron spin exchange within donor band (straight arrows).
    }
\label{fig:fig1}
\end{figure*}

While these general principles are well established, direct experimental determination of $T_2^*$ and $T_1$  at low fields, where they are crucially important for spin dynamics, is not straightforward, in particular for  {QW} electrons. 
Indeed,  even nominally undoped QWs host resident electrons with different degree of localization, and thus potentially different spin dynamics: donor-bound electrons, resulting from unavoidably present neutral donors (D0), and electrons localized on disorder potential, resulting from residual impurities and charge redistribution on the QW surface \cite{kosarev_microscopic_2019}. 
These different types of resident carriers are sketched in Fig.~\ref{fig:fig1}~(a). 
%
The orbits of donor-bound and disorder-localized electrons are shown in blue and magenta, respectively. 
The corresponding optical resonances, donor-bound exciton and negative trion are energetically close. 
In CdTe QWs they are separated by less than $1$~meV, close to their inhomogeneous broadening, Fig.~\ref{fig:fig1}~(b) \cite{Sergeev2005}. %
Therefore,  the spin properties of the two spin species can be hardly addressed independently using time-resolved techniques, such as polarized photoluminescence \cite{OO1} and pump-probe Kerr/Faraday rotation \cite{kikkawa_all-optical_2000,zhukov_spin_2009}, since they are based on picosecond pulses \footnote{It has been shown recently that it is it is possible to selectively address electrons bound to donors (stronger localization) or electrons localized on potential fluctuations (weaker localization) using spin-dependent photon echoes in combination with pump-probe Faraday/Kerr rotation~\cite{kosarev_microscopic_2019}}. 
%

%
%
By contrast, electron spin noise (SN) spectroscopy is a method of choice to achieve this goal. 
SN spectroscopy is based on a spectrally narrow continuous wave optical probe, that maps stochastic fluctuations of electron {spins} at thermal equilibrium onto Kerr (Faraday) rotation and/or ellipticity fluctuations of the probe interfering with the spin-flip Raman scattered light~\cite{aleksandrov81,Smirnov-review}. 
Analyzing the corresponding Fourier spectrum {makes} it possible to access the electron spin dynamics near equilibrium with minimal parasitic effects induced by photoexcited carriers. 
This technique, originally developed for atomic systems \cite{aleksandrov81,Crooker2004}, has been successfully applied for studies of electron spin properties in bulk semiconductors, such as GaAs \cite{oestreich_spin_2005} and, more recently, in CdTe \cite{Gribakin2025}, but only rarely in QWs \cite{muller_spin_2008,noise-trions,Ryzhov2015}. 

In this {work,} we show that analyzing the SN spectrum of donor-bound electrons in unintentionally doped CdTe QW within a relevant model makes it possible to identify the mechanisms of spin dephasing and relaxation.
First of all, the spin dephasing is essentially due to the distribution of hyperfine fields experienced by localized electrons \cite{merkulov02}. The so-called "satellite" peak, which corresponds to electron spin precession in random nuclear fields is very well defined in low-temperature SN spectra, see Fig.~\ref{fig:spectrum_fit}. Its FWHM allows us to determine the corresponding spin dephasing time, $T_2^*\approx 10$~ns.

Second, it appears that the electrons contributing to the SN spectrum are only those in the spectrally narrow subband of D0X transitions resonant with the probe laser. 
This selects only a small fraction of electrons within a large distribution of energies induced by the QW disorder.
As a result, even very brief (compared to expected electron spin-orbit relaxation time in the conduction band)  jumps  of the electron spin into the conduction band  followed by its relocalization on a donor with a slightly different energy within this distribution leads to the loss of the spin at the energy defined by the probe, see Fig.~\ref{fig:fig1}~(c). 
We show that this {spectral diffusion in the donor band, which is expected to be large in QWs, effectively destroys the temperature-activated slowing down of the nuclei-induced spin dephasing} observed in bulk CdTe and GaAs \cite{Croneberger2021,Sterin2022,Gribakin2025}.

The paper is organized as follows. The next section describes the QW sample and the experimental setup. In Section~\ref{sec:experiment}, main experimental results are presented: SN spectra measured at different probe {wavelengths and powers}, as well as at different temperatures. In Section \ref{sec:model}, we describe {a} theoretical model which allows us to explain the observations and to extract the parameters of the spin dynamics. Section~\ref{sec:concl} contains general summary and conclusions.

\section{Sample and experimental technique}
\label{sec:sample}
We study a CdTe QW of the width $L=9$~nm,  sandwiched between Cd$_{0.40}$Mg$_{0.60}$Te barriers.  The sample is grown by molecular beam epitaxy on a $[100]$ Cd$_{0.96}$Zn$_{0.04}$Te substrate. The details of the epitaxial structure are given in Table~\ref{tab:sample}. Note, that an additional thin QW close to the surface is intended to collect the holes likely to come from the surface states~\cite{PhysRevB.83.235317}. 

The photoluminescence (PL) and reflectivity of the sample measured at $T=5$~K are shown in Fig.~\ref{fig:fig1}~(b). 
The main PL peak can be  attributed to donor-bound excitons, and the peak at its high-energy shoulder to negative trions {localized at the fluctuations of the confining potential}. This interpretation is corroborated by the {SN measurements} presented in the next section. These two peaks dominate QW emission, while the free exciton resonance, expected $\approx 2.5$~meV above the trion, is not very well defined. 

\begin{table}[h!]
\centering
\begin{tabular}{|c| c| c|}
\hline
\textbf{Material} & \textbf{Mg, Mn or Zn content} & \textbf{Thickness, nm}\\ \hline
CdZnTe & 4  \% &  Substrate           \\ \hline
CdMnTe & 25  \%&   Buffer           \\ \hline
CdMgTe & 60  \%& 77  \\ \hline
CdMnTe & 25  \%& 68        \\ \hline
CdMgTe & 60  \%& 225   \\ \hline
\bf{CdTe} & \bf{0}  &    \bf{9 (QW)}          \\ \hline
CdMgTe & 60  \% & 148  \\ \hline
CdMnTe & 25  \%& 66         \\ \hline
CdTe & 0  &    4 (QW)         \\ \hline
CdMnTe & 25  \%& 32         \\ \hline
\end{tabular}
\caption{Structure of the sample. The studied QW layer is highlighted in bold.  }
\label{tab:sample}
\end{table}

The sample is placed in a variable temperature closed-cycle cryostat surrounded by a Helmholtz coil. The resulting magnetic field is oriented in the plane of the sample.

\begin{figure}
\includegraphics[width=3.5in]{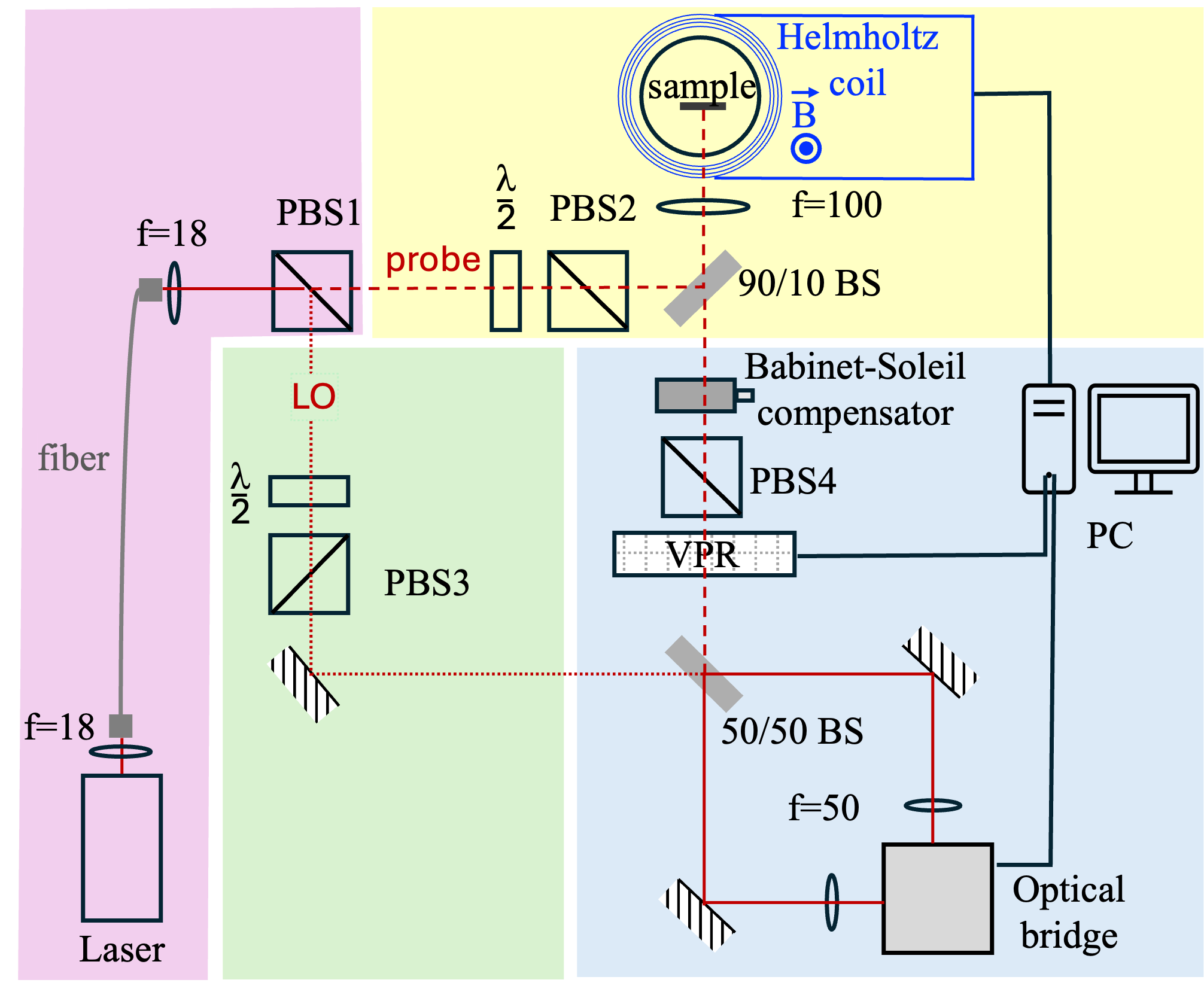}
\caption{Sketch of the experimental setup. Four regions identified by colour {show the laser excitation part (magenta), the split LO and probe beam paths (green and yellow, respectively), and the detection part (blue). The latter features optical mixing of the spin-flip Raman scattered light and the LO and its detection using the optical bridge}. Legend: PBS $\equiv$ Polarizing beam splitter; BS $\equiv$ Beam splitter; VPR $\equiv$  Variable phase retarder; $\lambda/2$ $\equiv$ Half-wave plate. Focal lengths are given in millimetres. 
    }
\label{fig:setup}
\end{figure}

The experimental setup used in this work is presented in Fig.~\ref{fig:setup}. 
The experiments are based on the homodyne detection of spin noise, which allows us to work with very low power of the probe {beam}, down to $<100$~µW. This approach has been successfully implemented in various configurations~\cite{RSI2016,Petrov2018,Sterin2018,Kamenskii2020,Cronenberger2019,Gribakin2025}. We will see below, that the low excitation power is crucial for the reliable determination of the spin properties.
%

The excitation part of the setup (magenta) starts from a continuous-wave, low-noise  Ti:Sapphire laser, tunable around QW optical resonances.  The laser beam  is sent through a mono-mode, polarization-conserving  optical fiber onto a polarizing beam splitter (PBS). The PBS1 splits the laser beam into two parts: the probe beam (dashed line), and the local oscillator (LO, dotted line). The power of the beams is then independently adjustable via dedicated half-wave plates and polarizers.  A beam sampler directs a small part ($<$10~\%) of the horizontally polarized probe normally to sample surface. The beam is focused by the $100$~mm focal-length  lens, yielding a Gaussian spot with a full width at half maximum {(FWHM)} of $w=21$~µm on the sample surface.

 The SN signal is contained in the spin-flip scattered light, which is emitted in the same direction as the reflected probe, but its polarization is vertical, while the probe beam is polarized horizontally.

 In order to detect the corresponding signal (blue part of the setup), the reflected probe beam is filtered out with $1/1000$ extinction ratio by a Babinet-Soleil compensator and PBS4, while the scattered light is transmitted and passes through a variable phase retarder (VPR).
 Its neutral axis is rotated at $\pi/4$ with respect to the horizontal axis and its retardance is  {set} either  {to} 0 (Measurement 1) or $\pi/2$ (Measurement 2).
 Then this light is mixed with the LO in a $50/50$ beam splitter and sent to the optical bridge, which converts  differential current fluctuations into voltage fluctuations. Then, to convert the voltage fluctuations into the noise power spectrum, an AC240 acquisition card performs a Fourier transform of the digitized time-dependent voltage. The SN spectrum is obtained by {subtracting} the spectrum of Measurement 1 (which corresponds to the shot noise) from that of Measurement 2 (which contains the SN signal) \cite{Gribakin2025}. 
 
Note that, since the rapidly fluctuating optical phase between probe and LO  arms of the interferometer is not stabilised, the differential signal on the optical bridge contains both ellipticity and Kerr rotation noise contributions \cite{Petrov2018}. 
However, we have checked that in our system the ellipticity contribution is negligible, as also observed in Ref.~\onlinecite{Petrov2018}. 
Therefore, the total SN power, that averages out optical phase fluctuations,  corresponds to $1/2$ of the Kerr rotation-induced SN power. 

Finally, to get rid of any spurious non-magnetic contributions, in this work  we analyse only the differences between the SN spectra measured in the absence and in the presence of the in-plane  magnetic field.

\begin{figure}
\includegraphics[width=3.5in]{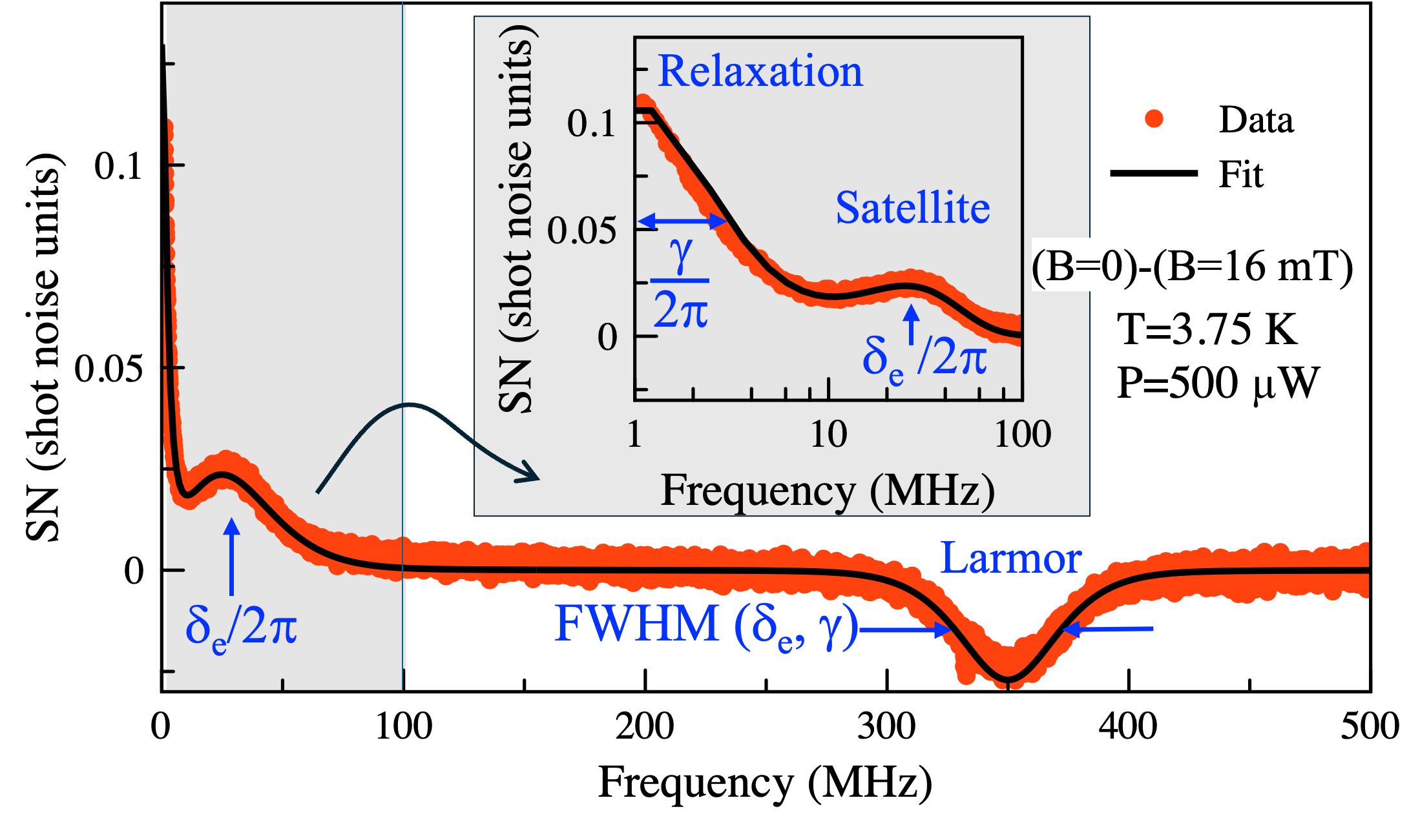}
\caption{SN {spectrum} measured at $B=0$, from which the spectrum measured under identical conditions at $B=16$~mT is substracted. Three main {features} are identified in this differential spectrum: the relaxational peak characterized by the width $\gamma/2\pi=1/T_1$, the satellite peak centred at frequency $\bar{\delta}_e/2\pi$, and the peak at Larmor frequency with the width depending on both $\bar{\delta}_e$ and $\gamma$. Inset shows low-frequency region in logarithmic scale. Red symbols show the experimental data, black solid lines are the results of the fit within the model of Section \ref{sec:model}.
Probe energy $E_p=1.616$~eV.    }
\label{fig:spectrum_fit}
\end{figure}

\section{Experimental results}
\label{sec:experiment}
\subsection{Main features of a differential SN {spectrum}}
\label{subsec:main_features}
Figure~\ref{fig:spectrum_fit} shows a typical differential SN spectrum measured with the probe energy tuned at $E_p=1.616$~eV [see Fig.~\ref{fig:fig1}(c)], slightly below the D0X resonance, with power $P=0.5$~mW.
The measurements are taken at $T=3.75$~K, and the difference between the SN spectra measured at $B=0$ and at $B=16$~mT is presented. 
%

The spectrum shows three main features. 
At approximately $F_L= 350$~MHz the negative peak results from electron spin precession at Larmor frequency $\Omega_L{=2\pi F_L}$ around magnetic field $B=16$~mT. {This frequency is given by $\hbar\Omega_L=\gamma_eB=g_e\mu_BB$, where} $\gamma_e$, $g_e$ and $\mu_B$ are electron gyromagnetic ratio, electron $g$-factor and Bohr magneton, respectively.
The linear fit of the Larmor frequency measured at various magnetic fields yields $g_e=1.54$, $\gamma_e=2.157$~MHz/G, consistent with the data available for electrons in CdTe QWs \cite{zhukov_spin_2009,Garcia-Arellano2019}. 
The width of the Larmor peak depends on both spin relaxation and coherence times, as will be discussed below.

The positive contribution at low frequencies is related to the SN at zero field. 
It consists of two peaks, one is centred at zero frequency and the other at $27$~MHz. The latter is referred to as the {``satellite''} peak. 
These {two} peaks correspond to the electron spin relaxation and {dephasing, respectively}~\cite{Glazov_Ivchenko_2012}. 
Physically, the electron spin dynamics consists of the precession in the random but quasi-static nuclear field and a slow relaxation of the spin component parallel to this field ~\cite{merkulov02}. 
The half-width at half-maximum (HWHM) of the zero-frequency peak corresponds to the relaxation time $T_1$.
The position of the ``satelite'' peak at $27$~MHz determines the typical electron spin precession frequency in a random nuclear field ${\bar\delta_e}/2\pi$ {($\bar{\delta}_e$ is the average value of $\delta_e$ in the ensemble, see below)}. 
This frequency matches the one observed in bulk CdTe samples,   suggesting that in our $9$-nm-wide QW, the SN signal is mainly contributed by bulk-like donor-bound electrons.
The satellite peak frequency can also be linked to the spin dephasing time as $T_2^*=\sqrt{\pi}/\bar{\delta}_e\approx$10~ns~\cite{Gribakin2025}.
%

\subsection{Probe energy and power {dependencies}}
\label{subsec:probe_energy}

\begin{figure}
\includegraphics[width=3.5in]{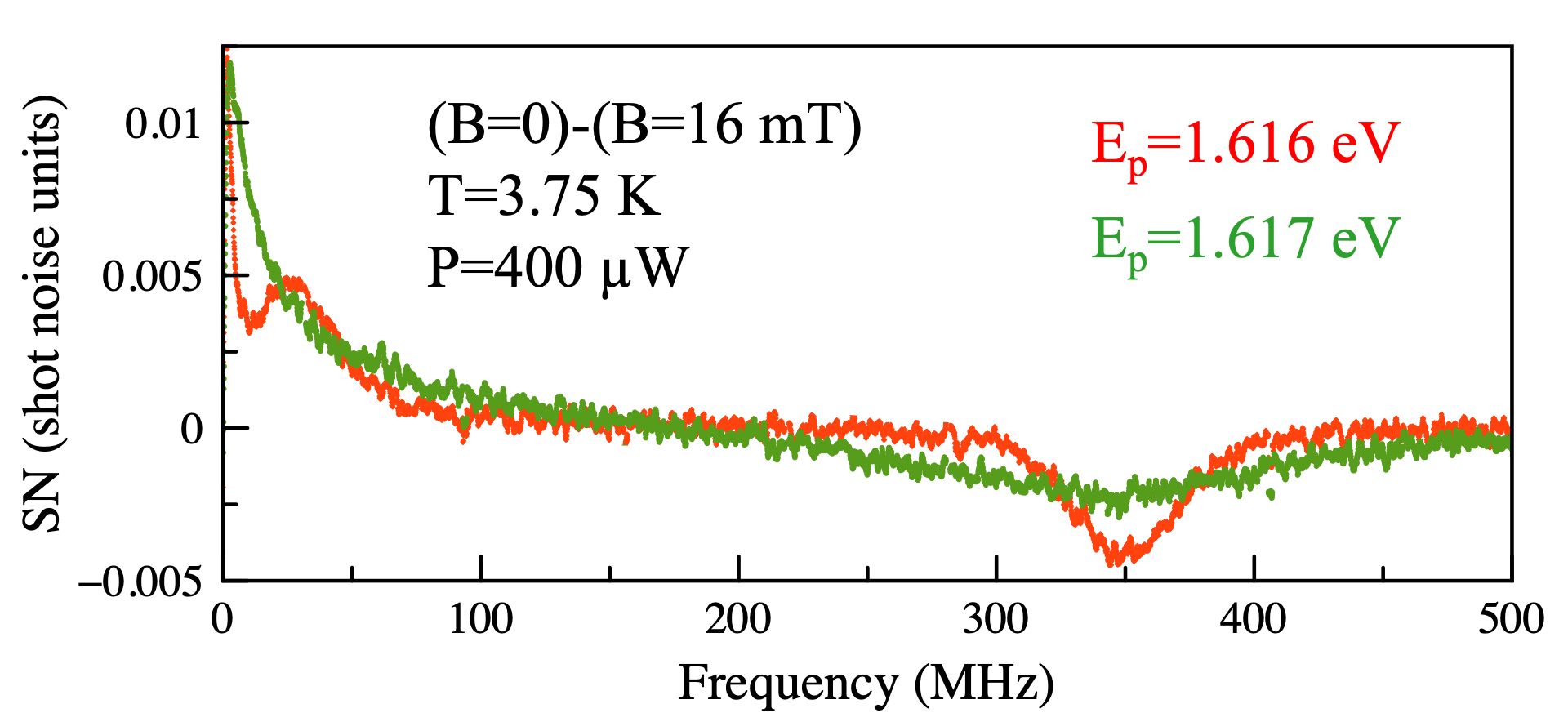}
\caption{ Differential SN spectra measured at two different probe energies, $E_p=1.616$~eV and $E_p=1.617$~eV. $P=400$~µW, $T=3.75$~K. At higher energy relaxational,  satellite and Larmor  peaks broaden significantly.
    }
\label{fig:Fig_WL_spectra}
\end{figure}

It is important to note that the {satellite} peak has rarely been observed so {clearly~}\cite{berski_interplay_2015,Gribakin2025}. 
To understand the reasons for that, let us compare the SN spectra measured under the same conditions but at slightly different probe energies, see 
Fig.~\ref{fig:Fig_WL_spectra}. 
It appears that the differential spectra measured at probe energies separated by only $1$~meV are drastically different. 
Indeed, both Larmor and satellite peaks are significantly {broader} at the higher energy. 
{This contribution cannot be separated from the relaxational part of the signal in a straightforward way, which makes the quantitative analysis of the zero-field part of the SN spectrum impossible without further considerations.} 

\begin{figure}
\includegraphics[width=3.5in]{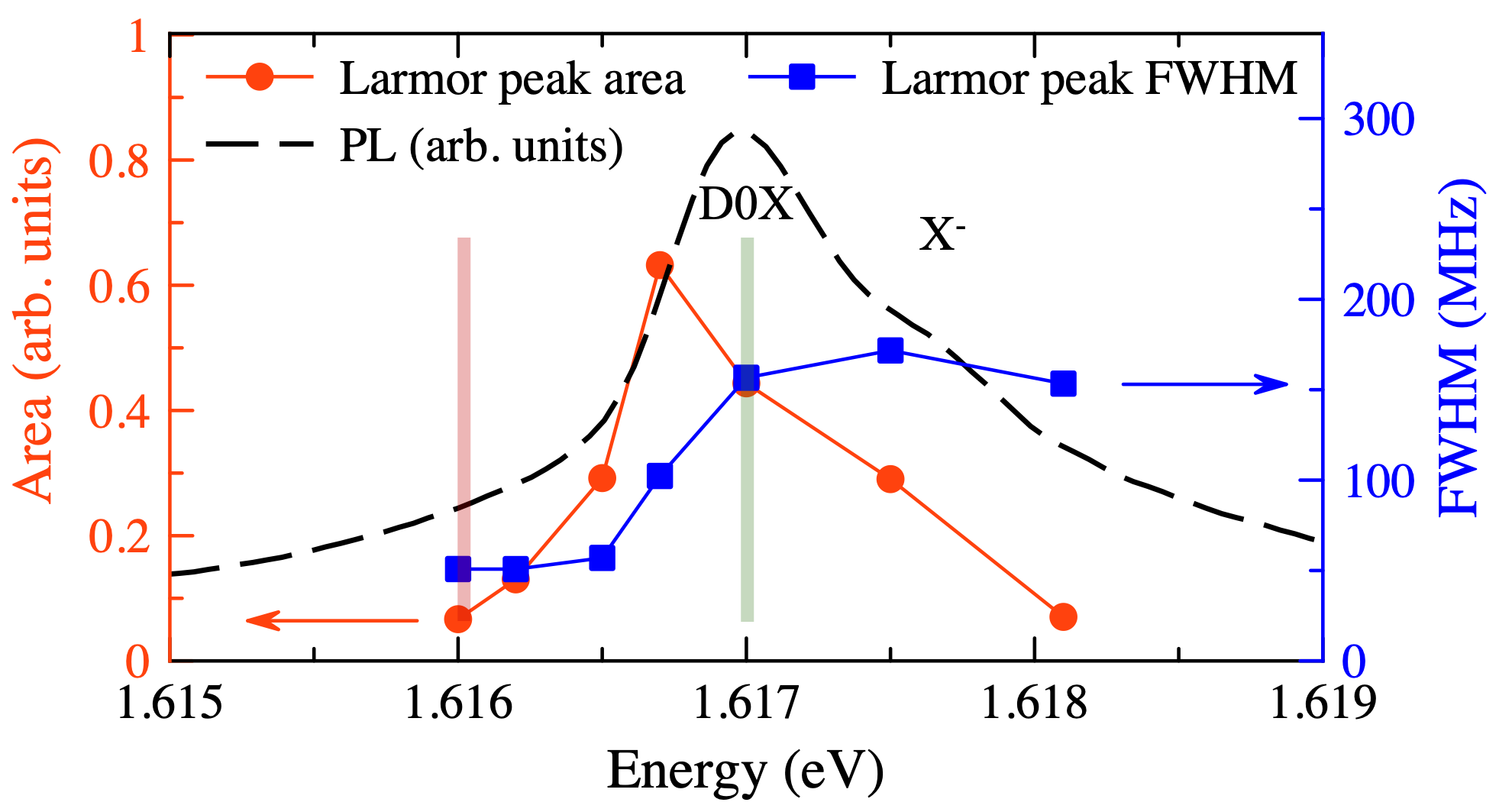}
\caption{Probe energy dependence of the SN: Area of the Larmor peak (brown circles, left {axis}) and its FWHM (blue squares, right  {axis}) as a  {functions} of the probe energy. PL spectrum is also shown for easier identification of the optical resonances. 
Green and pink vertical lines point the probe energies corresponding to the spectra shown in Fig.~\ref{fig:Fig_WL_spectra}.
    }
\label{fig:Fig_WL}
\end{figure}

\begin{figure}
\includegraphics[width=3.5in]{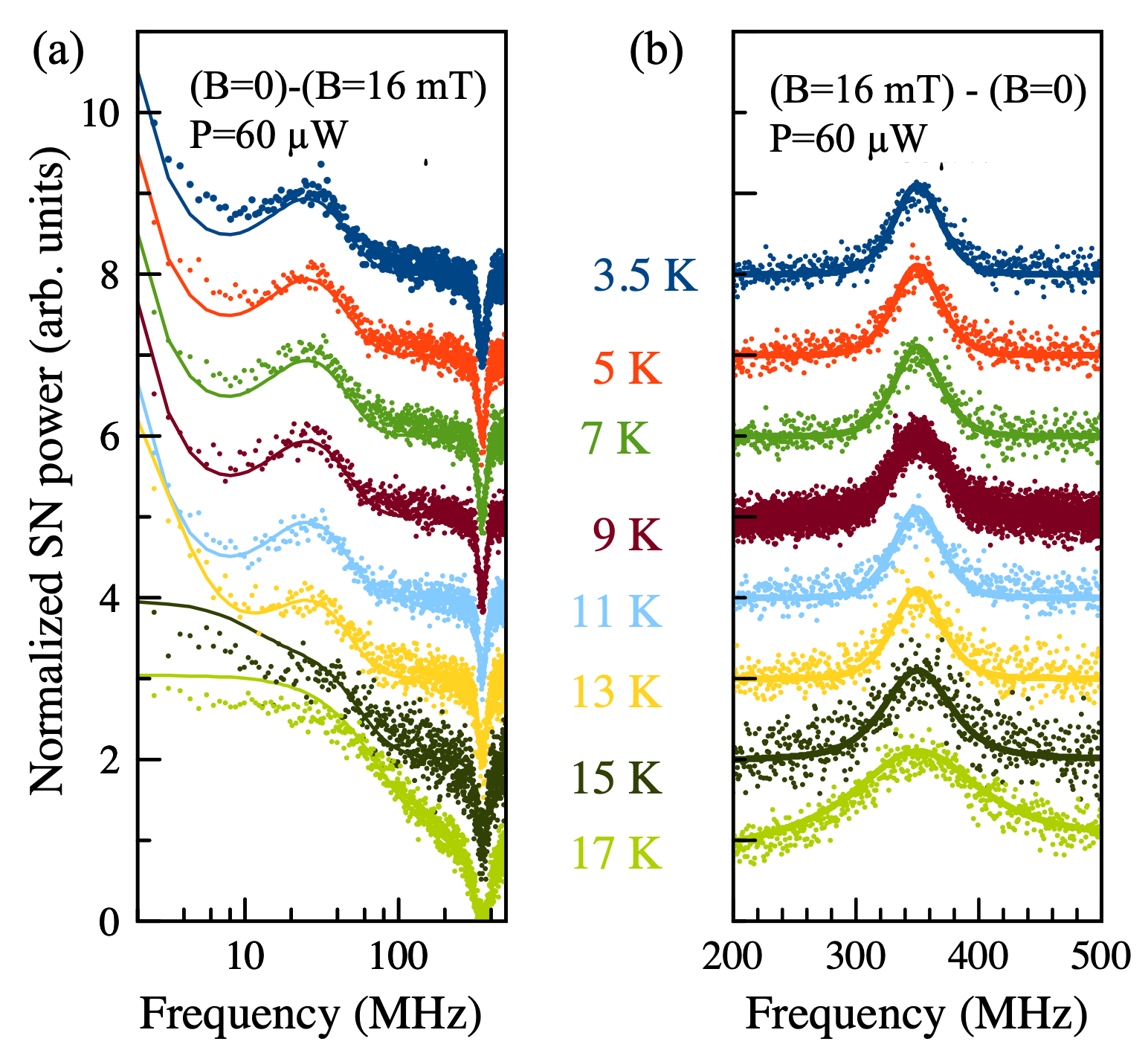}
\caption{Waterfall {plots} of the differential SN spectra measured at $P=60$~µW, $E_P=1.616$~eV and temperature varying from $T=3.5$~K to $17$~K. (a) Difference between the spectra measured  at $B=0$ and $B=16$~mT, frequencies in log scale, (b) Same as (a), but with opposite sign and with linear frequency scale. Symbols stand for the experimental data, solid lines are fits within the model of Eqs.~\eqref{eq:Sz2} and~\eqref{eq:Sz2_B} with the parameters given in the text. All the spectra are normalized to their Larmor peak amplitude.
    }
\label{fig:TdepSpectra}
\end{figure}
A systematic analysis of the probe energy dependence of the Larmor peak area and width is presented in Fig.~\ref{fig:Fig_WL}. 
Green and pink vertical lines indicate the probe energies corresponding to the spectra shown in Fig.~\ref{fig:Fig_WL_spectra}, the PL spectrum is shown on top for comparison. 
One can see that the SN signal area is approximately resonant with D0X - X$^-$ energy and has a clear peak-like shape. 
The FWHM of the Larmor peak increases from $20$ to $150$~MHz with increasing probe energy. 
By measuring SN spectra at different magnetic fields, we have checked that within experimental precision, the $g$-factor does not  present any measurable dependence on the probe energy.

%
These results can be tentatively interpreted as a signature of an increasingly important contribution to the SN signal of the resident electrons which are not bound to donors, but rather localized on the fluctuations of the confining potential. 
The latter can be due to non-uniformity of the Mg distribution in the QW barriers. 
In what follows, we focus on the study of donor-bound electrons, and keep the probe energy as low as possible to reduce the absorption and ensure the minimal contribution of other types of quasi-particles.


To access the intrinsic values of spin life- and coherence times, it is essential not only to reduce optical absorption, but also to minimize the excitation power.
We have checked that the SN spectra measured $E_p=1.616$~eV and at probe powers below $200$~µW have an identical shape.
The integrated SN power within the Larmor peak,  $\mathcal S$, expressed in shot noise units, increases linearly  with the probe power,
\begin{equation}
\mathcal S=\beta P,
\label{eq:Power}
\end{equation}
where $\beta/2\pi=1.5$~MHz/mW.

At  powers higher than $300$~µW, the spectra broaden significantly. 
%
%
%
%
%
This could be a consequence of the faster electron spin relaxation in the excited state or an uncontrolled heating of the sample.
Thus, to avoid these complications, the power is fixed at $60$~µW in temperature dependence measurements presented below.   

\subsection{Temperature dependence}
\label{subsec:temp}

SN dependence on the sample temperature is shown in Fig.~\ref{fig:TdepSpectra}. 
The difference between the spectra measured  at $B=0$ and $B=16$~mT is shown in Fig.~\ref{fig:TdepSpectra}(a) {on a linear-log} scale. The zoom  into the  Larmor peak part of the spectrum is shown in Fig.~\ref{fig:TdepSpectra}(b), with the opposite sign and {on a linear} scale.
The spectra are normalized to their Larmor peak intensity.
With increasing temperature, the Larmor peak broadens, while zero frequency and satellite peaks  merge. 
This behaviour is quite different from that observed in bulk CdTe, where an increase in temperature {from $5$ to} $15$~K leads to the so-called motional narrowing of the spin resonance.
In the next section we present a theoretical {model which} allows us (i) to understand this difference between SN spectra in bulk and QWs, (ii) to describe the shape {of} the spectra and determine the {electron} spin dephasing and spin relaxation {times,} (iii) to estimate the resident electron density.


%

\section{Model and comparison with experiments}
\label{sec:model}
\subsection{{Model of SN}}
\label{subsec:SN model}

{A basic} model of the spin noise of the localized electrons {was developed} in Ref.~\onlinecite{Glazov_Ivchenko_2012}. Following this work, we assume that the electron spin dynamics is described by the Bloch equation
\begin{equation}
  \frac{\d\bm S}{\d t}=(\bm\Omega_N+\bm\Omega_L)\times\bm S-\gamma\bm S,
\end{equation}
where $\bm\Omega_L$ ($\bm\Omega_N$) is the {Larmor} spin precession frequency corresponding to {external (nuclear)} field, and
$\gamma=1/T_1$ is {the longitudinal} spin relaxation rate unrelated to the hyperfine interaction.
To describe it, we note that the typical homogeneous linewidth of the optical transitions $\sim1~\mu$eV is three orders of magnitude smaller than the inhomogeneous broadening of the D0X {resonance} $\sim1$~meV. Therefore,  only a tiny fraction of localized electrons contributes to the SN signal at a given probe light energy, $E_P$. As a result, any {jump} of electron or of electron spin to another donor removes its contribution from the probed spin ensemble. In what follows, we assume that the effective spin relaxation rate $\gamma$ is determined by this type of processes. This is a specific feature of the distribution of the D0X energies, characteristic {of} QWs, presumably due to a broader distribution of D0X energies. It is not relevant to the SN spectra in bulk semiconductors~\cite{Gribakin2025}.

The dephasing of the electron spin is related to the hyperfine interaction. {We use a semi-classical {description} of electron spin precession in a field of ``frozen'' nuclear spin fluctuations. The} distribution of the corresponding frequencies is Gaussian:
\begin{equation}
  {\mathcal F_{\delta_e}}(\bm\Omega_N)=\frac{\e^{-\Omega_N^2/\delta_e^2}}{(\sqrt{\pi}\delta_e)^3}.
\end{equation}
The frequency $\delta_e$ is given by~\cite{merkulov02,book_Glazov}:
\begin{equation}
     \delta_e=\frac{1}{\hbar}\left(\frac{2}{3}I (I+1) \frac{ v_c}{V_e} \sum_j(A_j^2 x_j) \right )^{1/2},
    \label{eq:deltae}
\end{equation}
where $v_c=68$~${\mathring{\text{A}}}^3$ is the CdTe primitive cell volume, $I=1/2$ is the nuclear spin in CdTe, $x_j$ are the natural abundances of the three magnetic isotopes $^{111}$Cd, $^{113}$Cd, $^{123}$Te, and $A_j$ are the corresponding hyperfine interaction constants. Using the values for CdTe reported in Ref.~\onlinecite{Zhukov2014},
we obtain:
\begin{equation}
     \left({ \sum_j(A_j^2 x_j)}\right)^{1/2}=\mathrm{23}\ \textrm{µ}\mathrm{eV}.
\end{equation}
  
%

{The electron localization volume} for bulk donor-bound electrons is $V_e=8\pi a_B^3$, where $a_B$ is the Bohr radius, $a_B=4.9$~nm in CdTe  \cite{Garcia-Arellano2019,Marple1963}. {Its substitution to Eq.~\eqref{eq:deltae} gives $\delta_e/2 \pi=19$~MHz. 


The volume of the electron localization in a QW may, however,  depend on the position of the donor relative to the interfaces. This leads to the distribution of $\delta_e$ in the ensemble. We account for this by a convolution of} $\mathcal F{_{\delta_e}}(\bm\Omega_N)$ with a narrow Gaussian distribution of $\delta_e$ values~\cite{PhysRevLett.94.116601}.
This renormalizes the distribution of $\bm\Omega_N$ as:
\begin{equation}
  \tilde{\mathcal F}({\bm\Omega_N})
  =
  \int\frac{1}{\sqrt{\pi}\alpha{\bar{\delta}_e}}
  \exp\left[
  -\frac{{(\delta_e-\bar{\delta}_e)^2}}{{(\alpha\bar{\delta}_e)^2}}
\right]
{\mathcal F_{\delta_e}({{\bf\Omega}_{N})}}
  \d{\delta}_e.
\end{equation}
Here $\bar{\delta}_e$ is the average value of $\delta_e$, and the dimensionless parameter $\alpha<1$ describes its dispersion.

Under {the} above considerations, the SN spectrum in zero magnetic field takes the following {form}~\cite{Glazov_Ivchenko_2012,Smirnov-review}
\begin{multline}
  \label{eq:Sz2}
  (\delta S_z^2)_\omega\big|_{[B=0]}={\frac{\pi}{6}\int\limits_0^\infty\d\bm\Omega_N\tilde{\mathcal F}(\bf \Omega_N)}\\\times\left[\Delta(\omega)+{\Delta(\omega+\Omega_N)+\Delta(\omega-\Omega_N)}\right],
\end{multline}
where
\begin{equation}
  \Delta(x)=\frac{1}{\pi}\frac{\gamma}{x^2+\gamma^2}
\end{equation}
represents a Dirac delta function broadened by the effective spin relaxation rate $\gamma$.

In a sufficiently strong transverse magnetic field, such that $\Omega_L\gg{\bar{\delta}}_e$, the spectrum simply reads
\begin{equation}
  \label{eq:Sz2_B}
  (\delta S_z^2)_\omega\big|_{[B\neq 0]}=\frac{\pi}{4}\int\d{\bm\Omega_N\tilde{\mathcal F}}(\bm\Omega_N)\Delta(\omega-\Omega_L-\Omega_{N,B}),
\end{equation}
where $\Omega_{N,B}$ is the component of the nuclear field parallel to the applied magnetic field.
Eqs.~{\eqref{eq:Sz2}-\eqref{eq:Sz2_B}} constitute the necessary basis for the calculation of the differential SN spectra:
\begin{equation}
  \label{eq:Sz2_diff}
  (\delta S_z^2)_\omega\big|_{\rm diff}=(\delta S_z^2)_\omega\big|_{[B=0]}-(\delta S_z^2)_\omega\big|_{[B\neq 0]}.
\end{equation}

This model for the differential SN spectrum contains three fitting parameters: $\alpha$, $\gamma$ and   $\bar{\delta}_e$, since the Larmor frequency $\Omega_L$ is determined by the electron $g$-factor which we have measured independently ($g_e=1.54$, see Section~\ref{subsec:main_features}). 
{Equations~\eqref{eq:Sz2}-\eqref{eq:Sz2_diff}} show that, as anticipated in Section~\ref{subsec:main_features} and illustrated in Fig.~\ref{fig:spectrum_fit}, the frequency and the  width of the satellite peak {[last term in Eq.~\eqref{eq:Sz2}]}, as well as the width of the Larmor peak {[Eq.~\eqref{eq:Sz2_B}]} are determined by {$\bar{\delta}_e$}. 
Fitting to the experimental data yields $\bar{\delta}_e/2\pi=27$~MHz. This matches perfectly the value reported in bulk CdTe \cite{Cronenberger2019_arxiv,Croneberger2021,Gribakin2025} and suggests that, at least at the chosen probe energy, the SN signal is mainly contributed by the donor-bound electrons, which are only weakly affected by the QW interfaces.

However, the frequency obtained in both bulk and QW CdTe samples exceeds by a factor of $1.5$ the theoretically estimated value, $\delta_e/2\pi=19$~MHz in bulk CdTe, while the analogous calculation for GaAs is consistent with the corresponding experiments {\cite{berski_interplay_2015}}.
Because the calculations are based on the available  hyperfine constants and the value of the electron Bohr radius in bulk CdTe, we should conclude  that these {CdTe} parameters need to be revisited in future studies.

The dispersion of nuclear field in the ensemble of the donors, $\alpha$, slightly changes the {satellite peak shape}. The best agreement is achieved with $\alpha=0.4$.
The {effective} relaxation rate, $\gamma$, {equals to the HWHM} of the zero-frequency  peak and contributes to the width of the Larmor peak {[first term in Eq.~\eqref{eq:Sz2} and Eq.~\eqref{eq:Sz2_B}]}. For the spectrum shown in Fig.~\ref{fig:spectrum_fit}, at $T=3.75$~K and $P=500$~µW we obtain $\gamma/2\pi=1.5$~MHz.

\begin{figure}
\includegraphics[width=3.5in]{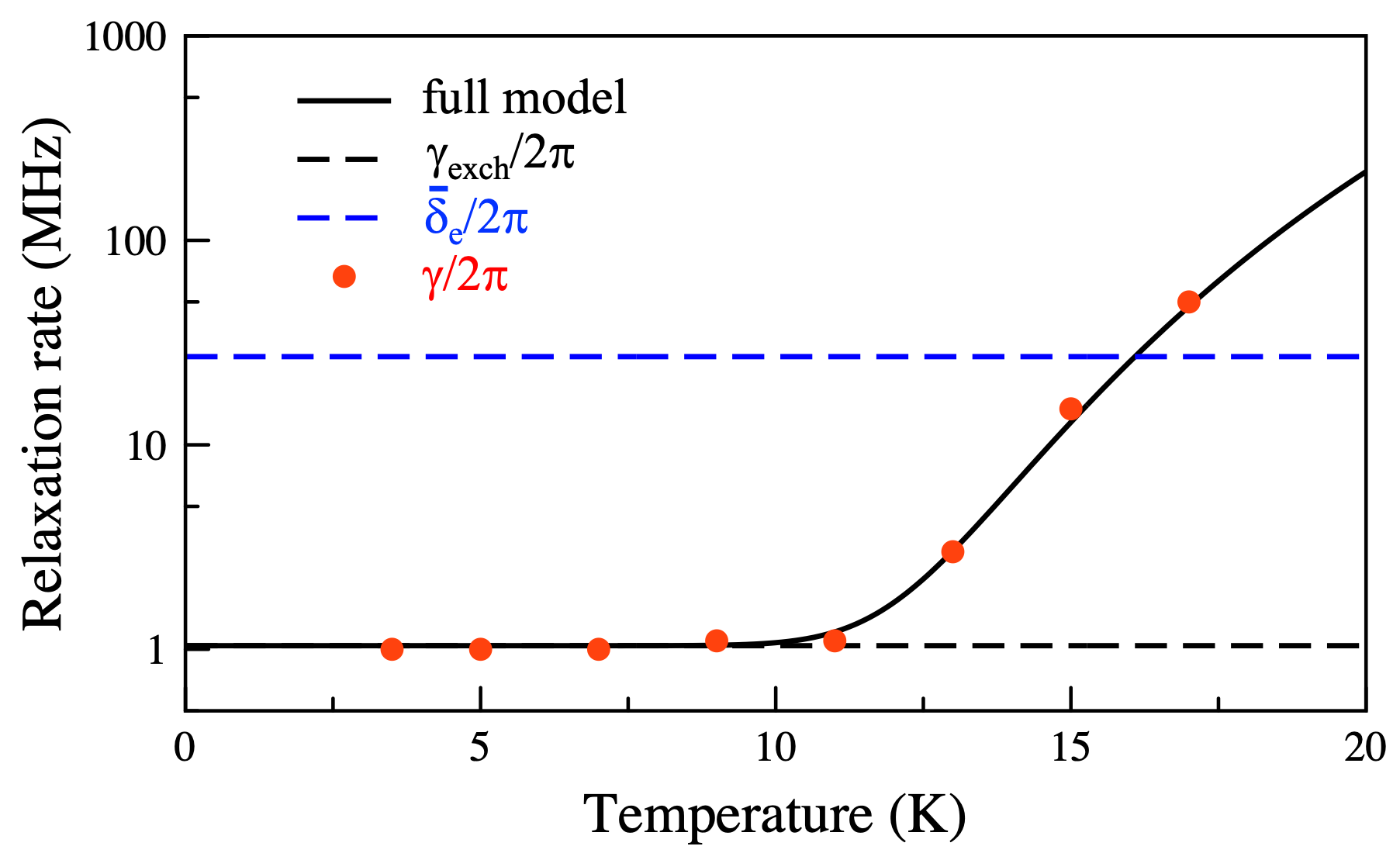}
\caption{Temperature dependence of {effective} electron spin relaxation rate $\gamma/2\pi$ extracted from the {fits shown} in Fig.~\ref{fig:TdepSpectra} (red dots). {The value} of $\bar{\delta}_e/2\pi=27$~MHz (kept fixed in the fits) is shown by blue dashed line. Black solid line is a fit {after Eqs.~\eqref{eq:gamma2} and~\eqref{eq:wa_T}} with $E_B=15$~meV, $C=5$~THz$\cdot\sqrt{\text{K}}$ and $\gamma_{\rm{exch}}/2\pi=1$~MHz (shown by black dashed line).
}
\label{fig:TdepFit}
\end{figure}

{To describe the temperature dependence of the SN spectra shown in Fig.~\ref{fig:TdepSpectra}, we note that} the distribution of the nuclear fields is {power-} and temperature-independent. Therefore, we keep the same values of $\bar{\delta}_e$ and $\alpha$, and use $\gamma$ as the only fitting parameter. The fits  of the experimental data using Eq.~\eqref{eq:Sz2_diff} are shown by solid lines in Fig.~\ref{fig:TdepSpectra}. The corresponding values of the {effective} spin relaxation rate $\gamma$ are shown in Fig.~\ref{fig:TdepFit}.

From Figs.~\ref{fig:TdepSpectra} and~\ref{fig:TdepFit} one can see that {the broadening of all {the} peaks in the SN spectra} with increase of the temperature is {well} described by the increase of the {effective} spin relaxation rate. {Merging of the zero frequency and satellite peaks at $T\gtrsim 15$~K corresponds to} the condition $\gamma\geq \bar{\delta}_e$ (blue dashed line in Fig.~\ref{fig:TdepFit}). {In the opposite limit of low temperatures,} the relaxation time $T_1=1/\gamma$ saturates at the value of $T_1=160$~ns ($\gamma/2\pi=1$~MHz), which {is among} the longest times observed in such structures in zero magnetic {field~\cite{Garcia-Arellano2019,PhysRevB.94.125401}}.

\subsection{Mechanisms of the spin relaxation}
\label{subsec:model_gamma}

To get an insight in the microscopic origin and the temperature dependence of the spin relaxation rate shown in Fig.~\ref{fig:TdepFit},  we separate it into two contributions:
\begin{equation}
  \label{eq:gamma2}
  \gamma=\gamma_{\rm{exch}}+w_{a}.
\end{equation}
The first one, $\gamma_{\rm{exch}}$, is the  spin exchange rate between nearest donors~\cite{KKavokin-review,garcia-arellano_exchange_2018}. It is temperature independent, and characterizes the spin diffusion in the impurity band. {As explained in Sec.~\ref{subsec:SN model}, in contrast with  bulk samples, the spin diffusion between donors acts  as effective spin relaxation, because it removes the given electron spin from the probed subensemble. Therefore, the exchange interaction without contribution from the spin-orbit coupling dominates the rate $\gamma_{\rm{exch}}$. The so-called anisotropic exchange interaction \cite{KKavokin-review} remains weaker than $\gamma_{\rm{exch}}$ for all temperatures.}

The second term in Eq.~\eqref{eq:gamma2}, $w_a$, is the temperature-dependent rate of the electron hopping to the conduction band~\cite{abakumov_1991}. After thermal activation, the electron is quickly captured back to the impurity band with the same spin, but it no longer contributes to the spin noise signal at the given probe wavelength. This process is {schematically shown by the yellow curved arrow in 
Fig.~\ref{fig:fig1}(c)}. 

The thermal activation rate depends exponentially on the donor binding energy $E_B$ and can be calculated using the detailed balance equation as follows~\cite{abakumov_1991,KKavokin-review}:
\begin{equation}
  \label{eq:wa}
  w_a=\frac{1}{2}N_cl_cv_t\exp(-E_B/k_BT).
\end{equation}
Here $N_c=mk_BT/(\pi\hbar^2)$ is the effective density of {electron} states in the first QW subband,  $m$ is the electron effective mass, $v_t=\sqrt{\pi k_BT/(2m)}$ is its  thermal velocity in the QW {plane, and} $l_c$ is the cross-section of the electron capture to the donor. 
The latter is quite difficult to calculate, since in polar semiconductors the dominant mechanism of the electron-phonon interaction is the piezoelectric interaction~\cite{birpikus_eng}, and thus the electron-phonon scattering matrix elements are polarization dependent~\cite{zook_PhysRev1964,polupanov_SovPhysSemicond1977,abakumov_SovPhysSemicond1977}. 
However, it has been shown that in bulk polar semiconductors {the capture} cross-section scales with the temperature as ${\sigma}_c^{\rm{bulk}} \propto T^{-4}$~\cite{abakumov_1991}. {We assume that in QWs, the capture length $l_c$ is of the order of $\sqrt{\sigma_c^{\rm{bulk}}}$}, so $l_c \propto 1/T^2$. This gives the temperature dependence of the activation rate in Eq.~\eqref{eq:wa} in the form
\begin{equation}
  \label{eq:wa_T}
  w_a=\frac{C}{\sqrt{T}}\exp(-E_B/k_BT),
\end{equation}
were $C$ is {a temperature-independent} constant.

Fitting {Eqs.~\eqref{eq:gamma2} and~\eqref{eq:wa_T}} to the temperature dependence of $\gamma$ {shown in} Fig.~\ref{fig:TdepFit} yields 
$\gamma_{\rm{exch}}/(2\pi)=1$~MHz, $C=5$~THz$\cdot\sqrt{\text{K}}$, and $E_B=15$~meV. 
While no theoretical values for $C$ are available in the literature and their calculation is quite complex, the obtained exciton binding energy is higher than the one expected for bulk donor-bound electrons, $\approx 11$~meV \cite{Garcia-Arellano2019}. %
%
%
This may be related to the effects of the electron size quantization in the QW.


For the estimation of the spin-exchange rate $\gamma_{\rm{exch}}/(2\pi)$ we rely on the calculations of Garcia-Arellano {\it et al.}~\cite{garcia-arellano_exchange_2018}.
This work establishes the relation between the spin exchange rate and the donor density. 
For $\gamma_{\rm{exch}}/(2\pi)=1$~MHz we get $N_d{\approx 8}\times10^{9}$~cm$^{-2}$. This value is consistent with our expectations for the unintentional doping in this structure. 
In the next {subsection}, we cross-check this consistency by analysing the intensity of the Kerr rotation {noise.}


\subsection{Estimation of the electron density}
\label{subsec:model_edensity}

In order to establish a relationship between the donor density and the measured signal, which is actually Kerr rotation noise expressed in the units of shot noise, we generalize the approach of Refs.~\onlinecite{Glazov_Ivchenko_2012,PhysRevB.85.195313} to the resonant homodyne detection scheme.  
Let us denote the LO polarization direction as $x$. Then the Kerr rotation signal $\mathcal K$ is defined as the difference between the intensities of the light components polarized along $x'$ and $y'$ axes, which {are tilted by} $\pm\pi/4$ {relative to} the $x$ axis:
\begin{equation}
  \mathcal K=P_{y'}-P_{x'}.
\end{equation}
Since the polarization rotation angle is small, it can be calculated as the following integral over the light beam cross section:
\begin{equation}
\label{eq:K1}
  \mathcal K=\frac{c}{\pi}\int\d\bm\rho\Re\left[E_y(\bm\rho){E_x^*}(\bm\rho)\right],
\end{equation}
where $E_x(\bm \rho)$ ($E_y(\bm \rho)$) is the amplitude of the $x$ ($y$) component of electric field in the detected beam, and $c$ is the speed of light. 
%
In the case of homodyne detection of Kerr rotation induced by localized electrons, {Eq.~\eqref{eq:K1}} can be rewritten as~\cite{yugova09,spherical-dots}:
\begin{equation}
\label{eq:K2}
  \mathcal K=-\frac{c{|t|^2}}{q^2\tau_0}\sum_j\Re\left[\frac{\mathcal E_P\left(\bm\rho_j\right)\mathcal E_{LO}^*\left(\bm\rho_j\right)\e^{\i\varphi}}{d_j+\i\Gamma_0}\right]S_{z,j}.
\end{equation}
Here $q=n\omega_P/c$ is the wave vector of the probe light with the angular frequency $\omega_P=E_p/\hbar$, $n$ is the refraction index of the sample, ${|t|^2}=4n/(n+1)^2$ is the intensity transmission coefficient through the sample surface, $\e^{\i\varphi}$ is a phase factor determined the sample thickness, $\tau_0$ and $\Gamma_0$ are the radiative lifetime and the homogeneous broadening of the D0X, respectively. 
The in-plane coordinate of the $j$-th donor is denoted as  $\bm \rho_j$, while $\d_j$ and $S_{z,j}$ stand for the corresponding angular frequency detuning and $z$-component of the spin. {$\mathcal E_P(\bm\rho)$ and $\mathcal E_{LO}(\bm\rho)$} are the electric field amplitudes of the probe and the LO {beams, respectively}. The two beams are assumed to overlap {perfectly and the sum runs over all donors.}
%

The spectrum of the Kerr rotation noise is related to the {SN spectrum} as
\begin{equation}
  (\mathcal K^2)_\omega=\left(\frac{c{|t|^2}}{q^2\tau_0}\right)^2\sum_j{\Re}^2\left[\frac{\mathcal E_P \left(\bm\rho_j\right) \mathcal E_{LO}^*\left(\bm\rho_j\right)\e^{\i\varphi}}{d_j+\i\Gamma_0}\right](\delta S_z^2)_\omega,
\end{equation}
where the spin noise spectrum, $(\delta S_z^2)_\omega$, is per electron, as above.
To calculate the sum in this expression, which will involve the donor density {$N_d$}, we assume an inhomogeneous broadening $\Gamma$ of the D0X resonance  around the central frequency ${\omega}_0$ in the  Gaussian form ${\propto N_d}\exp\left[-(\omega_j-\omega_0)^2/\Gamma^2\right]$. We also consider that  the power density distribution within  the light beam cross-section is Gaussian, that is  $\propto\exp(-\rho^2/a^2)$, where $a=w/(2\sqrt{\ln 2})$ with $w={21}$~µm being the experimentally determined FWHM of the light spot intensity on the sample surface. Assuming $\Gamma_0\ll\Gamma$, the noise intensity spectrum reads
\begin{equation}
  \label{eq:K2_ans}
  (\mathcal K^2)_\omega=\pi^{3/2}\left(\frac{{|t|^2}}{a q^2\tau_0}\right)^2\frac{P P_{LO}\e^{-(\omega_0-{\omega}_P)^2/\Gamma^2}}{\Gamma_0\Gamma}N_d(\delta S_z^2)_\omega,
\end{equation}
where $P$ and $P_{LO}$ are the powers of the probe beam and of the local oscillator, respectively.

Finally, {the} integrated noise intensity in the shot noise units {is given by}
\begin{equation}
  \mathcal S=\frac{1}{({I}_{\rm shot}^2)_\omega}\int_0^\infty(\mathcal K^2)_\omega\d\omega,
\end{equation}
were $({I}_{\rm shot}^2)_\omega=E_P{P}_{LO}$ is the {white shot noise} spectrum.
Substituting here Eq.~\eqref{eq:K2_ans} we obtain a linear relation between the {noise intensity} and the probe power, Eq.~\eqref{eq:Power}, which was checked experimentally. The experimentally measured factor $\beta$ is proportional to the donor density in the QW plane, $N_D$:
\begin{equation}
  \beta=\left(\frac{{|t|^2}}{aq^2\tau_0}\right)^2\frac{\pi^{3/2}\e^{-(\omega_0-{\omega}_P)^2/\Gamma^2}}{8E_p\Gamma_0\Gamma} N_d,
  \label{eq:beta}
\end{equation}

Thus, with experimentally determined $\beta/2\pi=1.5$~MHz/mW, $E_P=1.616$~eV and $w=21$~µm, estimated from PL spectrum $\hbar\Gamma=0.6$~meV, $\hbar\omega_0=1.617$~eV, and {taking} $n^2=7.1$ for CdTe~\cite{Madelung2004}, and $\Gamma_0=1/(2\tau_0)=10$~ns$^{-1}$~\cite{zhu07,PhysRevB.83.235317,PhysRevLett.109.157403,PhysRevB.96.075306}, we obtain  $N_d\approx10^{10}$~cm$^{-2}$ in reasonable agreement with the estimation based on the experimental value of the exchange interaction made in Sec.~\ref{subsec:model_gamma}.

\section{Summary and outlook}
\label{sec:concl}

We have measured and analysed theoretically the SN spectra of localized, donor-bound electrons in nominally undoped CdTe/CdMgTe QW. Until now, only SN spectra in the bulk CdTe have been studied and there are only few reports on SN spectroscopy in QWs, all based on GaAs \cite{muller_spin_2008,noise-trions,Ryzhov2015}.
Our results show that, in agreement with well-established theoretical models~\cite{Glazov_Ivchenko_2012}, at low temperatures and at zero magnetic field, electron spins precess in the hyperfine field induced by ``frozen'' fluctuations of the nuclear spins lying within the electron localization volume. This precession constitutes the main mechanism of electron spin dephasing. 
The corresponding peak in the SN spectrum, known as the ``satellite'' peak, appears at $27$~MHz at exactly the same frequency as in bulk CdTe, indicating that, at least in a QW of $9$~nm width, the volume of the donor-bound electron wavefunction remains  identical to the one in the bulk. 
By comparing this frequency  with the theoretical predictions based on known CdTe parameters, we show that the theory underestimates it by a factor of $1.5$. Since the calculations are based on the available hyperfine constants and the electron Bohr radius in bulk CdTe, we conclude that these parameters need to be revisited.
%

While in bulk CdTe and GaAs crystals the shape of the satellite peak has been described by the same dispersion of the nuclear field $\delta_e$ for all donors~\cite{Gribakin2025,berski_interplay_2015}, our data suggests that in a QW, the strength of the nuclear spin fluctuations varies significantly within the probed ensemble.
We account for this by introducing a supplemental parameter $\alpha$, characterizing the distribution of $\delta_e$ around the central frequency $\bar{\delta}_e$. It is found to be $\alpha \approx 40$\% of the mean nuclear spin fluctuation in our sample.

%

The application of a magnetic field in the plane of the sample confirms the zero-field results: the satellite peak transforms into the Larmor peak, but its shape, which at low temperatures and low electron densities is determined mainly by nuclear spin fluctuations, is consistent with that of the satellite peak. 
The corresponding dephasing time of the electron spin is very close to that measured in bulk CdTe, $T_2^*=\sqrt{\pi}/\bar{\delta}_e\approx10$~ns \cite{Cronenberger2019_arxiv,Croneberger2021,Gribakin2025}.

The satellite peak at $27$~MHz is so well-resolved that we are able to reliably separate the spin dephasing and {longitudinal spin relaxation} at zero magnetic field.  %
At the lowest temperatures {$T < 10$~K}, the relaxation time $T_1=1/\gamma=160$~ns.
By measuring the temperature dependence of the SN, we identify two contributions to effective spin relaxation: spin exchange between donors, which dominates at low temperature, and temperature-activated spin jumps to the conduction band. 
The spin exchange is related to the donor density. The estimation of this density based on the theory developed in Ref.~\onlinecite{garcia-arellano_exchange_2018} is consistent with the theoretical {estimate}  we make from the integrated SN power, $n_D\lesssim 10^{10}$~cm$^{-2}$. 
%

Finally, in contrast with bulk samples, either GaAs or CdTe, a very specific situation takes place in our QW, where the temperature-induced activation of electron hopping leads to the loss of the spin which we detect at the given probe energy, rather then to motional narrowing.
We argue that this is a consequence of donor energy dispersion in the impurity band, which is much broader than the detection band of the probe beam. As a result, even fast jumps into the band and back without loss of the spin polarization lead to the spectral diffusion of the spin outside the detection band.

As a future development, it is be important to understand the origin of the discrepancy between experimental and theoretical values of electron spin precession frequency in the  quasi-static nuclear field $\bar{\delta}_e$.
In order to {further} check the validity of the model, it would also be interesting to reduce the CdTe QW thickness down to the Bohr radius of the {bulk} donor-bound electron and below, to quantify the changes in the "satellite" peak frequency, and thus electron localization volume. We expect a blue shift that scales with the QW width as $a_B\sqrt{L}$  and a corresponding reduction of the spin coherence time. 
On the other hand, studies of donor-bound electrons in GaAs QWs in the same low-density regime can be performed. Depending on the spectral width of the impurity band in GaAs QWs, the temperature-induced activation of electron hopping  may lead either to motional narrowing and thus slowing down of electron spin relaxation, like in the bulk samples (narrow impurity band), or to the loss of spin via spectral diffusion as observed in this work. 


\section{Acknowledgements}

The authors acknowledge the Foundation for the Advancement of Theoretical Physics and Mathematics ``BASIS.''  R.A. has benefited from the technical and scientific environment of the CEA-CNRS joint team ``Nanophysics and Semiconductors''.  Financial support  from French National Research Agency, Grant No. ANR-21-CE30-0049 (CONUS) is gratefully acknowledged. The theoretical modelling by A.L.Z. and D.S.S. was supported by the Russian Science Foundation Grant No. 25-72-10031.


\renewcommand{\i}{\ifr}
\bibliography{refs}

\end{document}